\documentclass[12pt]{article}
\usepackage{epsf,epsfig,psfig,float,amssymb,latexsym}
\usepackage{graphics,psfrag}
\newcommand{\be}{\begin{equation}}
\newcommand{\ee}{\end{equation}}
\newcommand{\ba}{\begin{eqnarray}}
\newcommand{\ea}{\end{eqnarray}}

\newcommand{\nn}{\nonumber}

\newcommand{\vs}{\vspace{-0.20cm}}

\begin{document}

\thispagestyle{empty}

\vspace{2cm}

\begin{center}
{\Large{\bf The Mixing Angle of the Lightest Scalar Nonet}}
\end{center}
\vspace{.5cm}

\begin{center}
{\Large Jos\'e A. Oller}
\end{center}

\begin{center}
{\it {\it Departamento de F\'{\i}sica. Universidad de Murcia.\\ E-30071,
Murcia. Spain.\\
oller@um.es}}
\end{center}
\vspace{1cm}

\begin{abstract}
\noindent
We show that the $\kappa$, $a_0(980)$, $\sigma$ and the $f_0(980)$ 
resonances constitute the lightest scalar nonet in 
three different and complementary ways. First, by establishing the continuous movement of the poles 
from the physical to a SU(3) limit. Second, by performing an analysis of the couplings  of 
the scalar mesons to pairs of pseudoscalars and third, by analysing the
couplings of the scalars with meson-meson SU(3) scattering eigenstates. Every of  the last 
two methods agree that the mixing angle between the singlet and the octet
$I=0$ states is $\theta= 19^o\pm 5$ degrees, so that the $\sigma$ is mainly
the singlet and the $f_0(980)$ the isosinglet octet state. 
\end{abstract}

\vspace{2cm}


\newpage

\section{Introduction}
\label{sec:intro}
\def\theequation{\arabic{section}.\arabic{equation}}
\setcounter{equation}{0}

Just after the experimental discovery of the $f_0(980)$ \cite{esta73}, a few years after the one of the 
$a_0(980)$ \cite{a0},  Morgan in ref.\cite{morgan} made a first attempt to establish the states 
belonging to the $J^{PC}=0^{++}$ lightest scalar nonet and to calculate the
 mixing angle. He established, as a tentative solution, that the $f_0(980)$, $a_0(980)$ and 
a $\kappa(\backsim 1200)$ together with an extra $f_0(\backsim 1100)$ would
form this scalar nonet with $\cos^2\theta=0.13$, being $\theta$ the mixing angle. The $a_0(980)$ with $I=1$
and the $\kappa$ with $I=1/2$ are supposed to 
remain as pure octet states in this and in any other $SU(3)$ analysis that we refer here. In our
notation, $\cos\theta$ is the projection of the $f_0(980)$ on the isosinglet
octet state. The
result of Morgan's analysis is far from the ideal mixing, the one found in the vector and
tensor multiplets \cite{saku,oku,glashow,pic}, where $\cos^2\theta=2/3\simeq 0.67$, so
that the state coupling more strongly with strangeness is mainly an
octet. Notice that in the $\bar{q}q$ picture for mesons, the ideal mixing
implies that the heavier state is $\bar{s}s$ while the lighter is $(\bar{u}u+\bar{d}d)/\sqrt{2}$, 
the standard example being the $\phi$ and $\omega$ resonances.  Later on, Jaffe 
\cite{jaffe} developed his model based on the MIT Bag Model \cite{mit} for
multiquark hadrons as $\bar{q}^2 q^2$ states. 
In this calculation a $0^{++}$ nonet of scalar resonances emerges with rather light masses due to strong 
attractive gluon-magnetic interactions. The nonet is made up by a $f_0(\backsim 700)$, $f_0(980)$, $a_0(980)$, the latter two with predicted masses of 1100 MeV,  and a
$\kappa(900)$. In this model a natural explanation of the approximate
degeneracy of the $f_0(980)$ and $a_0(980)$ is obtained 
in terms of just their quark content, something that badly fails when these resonances are supposed to be 
$\bar{q}q$ states. Here, one ends with 
a dual situation with respect to the just referred ideal mixing, so that $\cos^2\theta=1/3=0.33$. 
 Another pioneering work on the scalar 
nonet is that of Scadron \cite{scad}, where making use of a quark model, he establishes as well that the 
lightest scalar nonet comprises the $\sigma(750)$, $\kappa(800)$, $f_0(980)$
and $a_0(980)$ so that $\cos^2\theta=0.39$.   
 More recently,  Napsuciale \cite{nap}, in the $U(3)\times U(3)$ linear sigma model with $U_A(1)$ symmetry 
breaking \cite{15,21,23}, predicted again that the lightest scalar nonet
consists of the $\sigma(\backsim 500)$, 
$f_0(980)$, $\kappa(\backsim 900)$ and the $a_0(980)$. In this calculation the properties of the pseudoscalar 
nonet $0^{-+}$ are fitted and those of the scalar nonet are predicted. In this
way, Napsuciale concludes that $\cos^2\theta=0.87$. It is worth mentioning
that both Napsuciale and Scadron use the so called strange and non-strange
basis, the one that results from ideal mixing, and they obtain the same
absolute value for the mixing angle in this basis, $|\phi|\simeq 15^0$, but differ in the
sign, so that $\theta=35.3+\phi$ is also different. Refs.\cite{tornq,hoft} also employ the same model of
ref.\cite{nap} but with rather different conclusions due to the strong sensitivity of
the results of this model in terms of the pseudoscalar experimental input, see
ref.\cite{nap} for a detailed discussion on this issue. Finally, Black et al. in
ref.\cite{putative}, from an analysis of $\pi\pi$ and 
$K\pi$ scattering from chiral invariant Lagrangians including bare resonances,
also consider the same resonances for 
the lightest scalar nonet with $\cos^2\theta=0.63$. There are several other groups that also
agree that the members of the lightest 
scalar resonances are the ones listed above, that is, $f_0(600)$ or $\sigma$,
$f_0(980)$, $\kappa(700-900)$ and $a_0(980)$
\cite{eef,ishida,iamprd,iam,nd,jamin,julich} and an accord is nowadays rising that
the lightest $\bar{q}q$ nonet, with rather standard properties, mixed 
with a glueball is made up by more massive resonances around the 1.4 GeV
region, the $K^*_0(1430)$, $a_0(1450)$, 
$f_0(1370)$, $f_0(1500)$ and $f_0(1710)$ \cite{largoclose,pdg}. From the
experimental point of view, new and precise 
experimental data has also helped to clarify to large extend the status of
the lightest scalar nonet. On the one hand, 
we have the $\phi$ radiative decays to $\gamma \pi\pi$ and 
$\gamma \pi\eta$ \cite{cloe,oller,rafel,acha,penni}, which rule out simple
$\bar{q}q$ models for the $f_0(980)$ and $a_0(980)$ resonances \cite{isgur,largoclose}, and on the other hand
 we have the $D^+$ decays, where the $\sigma$ and $\kappa$ resonances are clearly seen \cite{e791,gobel} with masses 
around 0.5 and 0.8 GeV and large widths, between 0.3--0.4 GeV. 

In this work, we take the strong S-wave T-matrices from refs.\cite{nd,npa} and analyse 
their spectroscopic content 
regarding the lightest scalar mesons. It is worth stressing that these T-matrices have been by now checked in a 
wide range of physical processes, namely, pure strong meson-meson scattering
\cite{nd,npa}, $\gamma\gamma$ fusion into two 
mesons \cite{gama}, $J/\Psi$ decays \cite{jpsi} and $\phi$ radiative decays
\cite{plb,oller}. In ref.\cite{nd} 
was already established that the $\sigma$, $f_0(980)$, 
$\kappa$ and $a_0(980)$ belong to the same scalar nonet but no attempt was made to study the mixing between 
the  singlet and the $I=0$ state of the octet making up the $\sigma$,
$f_0(980)$ resonances and to perform a SU(3) analysis 
of the couplings constants of the members of this lightest scalar nonet as similarly performed in the vector or 
tensor nonets \cite{saku,oku,glashow}. Here we aim to fill up this gap. We
perform two SU(3) analyses of the 
couplings constants of the scalar resonances by considering 
 in the first analysis  the couplings to two pseudoscalars isospin states and in the second one, 
the couplings to SU(3)
 eigenstates. The two methods have different sensitivity 
to the SU(3) breaking, follow very different numerical analyses, and their
agreement is taken as a guarantee of the consistency of our results. This is 
performed in section \ref{sec:form} where we also show the continuity of the
movement of the poles from the physical situation to a SU(3) symmetric one,
where the poles corresponding to the aforementioned scalar resonances 
bunch in a degenerate octet of resonances and in a lighter singlet pole. Some conclusions are presented in 
section \ref{sec:conc}.

\section{SU(3)-Analyses}
\label{sec:form}
\def\theequation{\arabic{section}.\arabic{equation}}
\setcounter{equation}{0}

\begin{figure}[H]
\psfrag{d00 (degrees)}{$\delta^{11}_{0}$ (degrees)}
\psfrag{dpiK (degrees)}{$\delta^{12}_{0}$ (degrees)}
\psfrag{ine}{$(1-\eta^2_{0})/4$}
\psfrag{dKpi (degrees)}{$\delta^{11}_{1/2}$ (degrees)}
\psfrag{despieta}{Events}
\psfrag{E (MeV)}{$\begin{array}{c} \\ \sqrt{s}\,\hbox{ MeV} \end{array}$}
\centerline{\epsfig{file=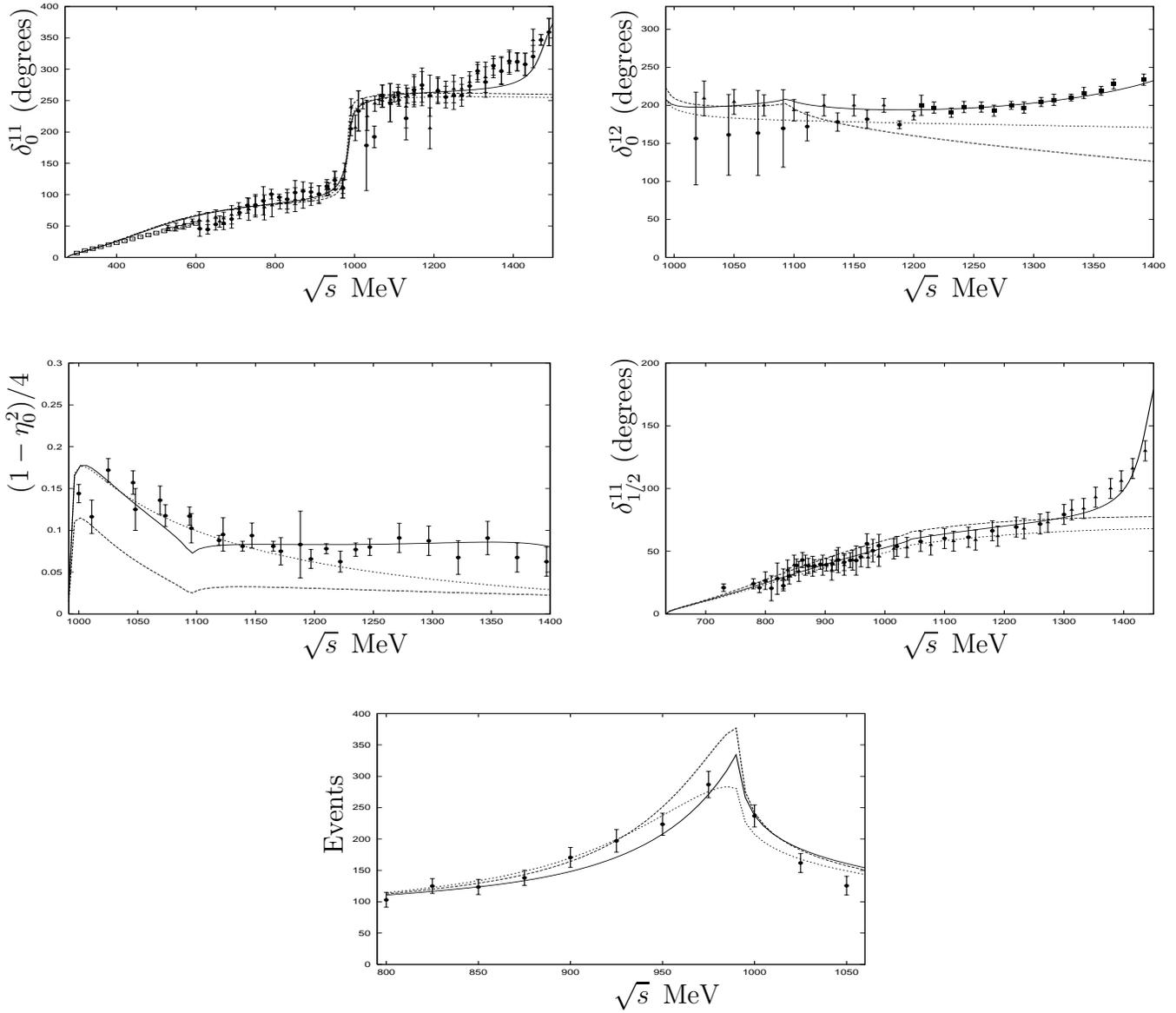,height=6.0in,width=7.0in,angle=0}}
\vspace{0.2cm}
\caption[pilf]{ \protect \small Fits to scattering data. The solid lines is the full approach of ref.\cite{nd}. 
The long-dashed lines correspond to the fit given in eq.(\ref{fita}) where the bare poles are removed and a common 
subtraction constant is used for all the channels. The short-dashed lines
correspond to remove both the bare 
poles and the $\eta_8\eta_8$ channel but allowing for different values of the subtraction constants, 
eq.(\ref{fitb}). From top to bottom and left to right, the first three
panels refer to the isosinglet S-wave sector: Elastic $\pi\pi$ phase shifts,
the inelastic $\pi\pi\rightarrow K\bar{K}$ phase shifts and $(1-\eta_0^2)/4$ with $\eta_0$
the inelasticity, in order. The fourth panel is the isodoblet S-wave $K\pi$
elastic phase shifts and the last one is an event distribution around the
isovector $a_0(980)$ resonance.   
\label{fig:1s0kswcm}}
\end{figure}

In table \ref{table:coup} we show the physical pole positions in the second Riemann 
sheet\footnote{The sheets of the complex $s$-plane are defined with respect to the branch cuts of the channel 
three-momenta $q_i$. The first or physical sheet corresponds to 
$\hbox{Im}q_i\geq 0$ for all the channels, and the second sheet corresponds to
changing the sign of the momenta of only the 
lightest channel, $\hbox{Im}q_1\leq 0$, while for the rest of channels
$\hbox{Im}q_i\geq 0$, $i\geq 2$.} 
from the T-matrices of refs.\cite{nd,npa}. These T-matrices are obtained once the chiral
 expansion is resummed to fulfill the unitarity cut to all
orders. The resummation is performed so that the chiral perturbative series
\cite{wein,gasleut} is recovered
algebraically at low energies when the chiral power counting holds. In
addition to refs.\cite{nd,npa},  a general discussion of the method is given in ref.\cite{nn}. 
In fig.\ref{fig:1s0kswcm} we show by
the solid lines the full results of ref.\cite{nd}, corresponding to the fit in
  eq.(73) of the latter reference. In this reference, the $\pi\pi$, $K\bar{K}$ and $\eta_8\eta_8$ states were 
included in the $I=0$ channel, the $K\pi$ and $K\eta_8$ in the $I=1/2$ one 
and the $\pi\eta_8$ and $K\bar{K}$ for $I=1$.
  
In ref.\cite{npa} a three-momentum cut-off was used to calculate the unitarity
two meson diagram, $g(s)$, but now we have calculated it  
through a dispersion relation so that 
we have a subtraction constant as free parameter for each channel. In the SU(3) limit, as shown 
in ref.\cite{twolamb}, all the subtraction constants must be equal. Taken this 
to be case,  we fit the common subtraction constant for all the isospin channels with the preferred 
value,
\be
\label{fita}
a(\mu)=-1.23, \hbox{ for } \mu=M_\rho ~,\\
\ee
where $\mu$ is the subtraction scale.  The formalism is invariant under changes of
the subtraction scale $\mu$ \cite{nd} such that $a(\mu')=a(\mu)+\log
\mu'^2/\mu^2$. In this fit we have also included the
$\eta_8\eta_8$ state for $I=0$ as in ref.\cite{nd}. The results are given by the long-dashed lines in
fig.\ref{fig:1s0kswcm}. Let us notice that the  differences with respect
to the full results of ref.\cite{nd}, solid lines, come from the fact that in
this fit we have removed the set of preexisting resonances of
ref.\cite{nd}. This comprises an octet of mass around 1.4 GeV and a singlet of
mass around 1 GeV. The role played by the heavier octet is apparent when
considering the figures in the region above 1.2 GeV. We see that the main
differences for lower energies correspond to the $I=0$ inelasticity,
third panel from top to bottom and left to right.

It is also interesting to remark that by performing a comparison between the
$g(s)$ function calculated with a cut-off 
or with the dispersive integral \cite{iamprd}, we can then obtain an
expression for the subtraction constant in 
terms of a natural-sized cut-off. This expression corresponds to:
\ba
\label{acutoff}
a(\mu)=-\log \frac{4 Q^2}{\mu^2}~.\nn
\ea
Thus, for $Q\simeq M_\rho\simeq 4\pi f_\pi$ GeV, we have
$a(M_\rho)$ in the interval $[-2,-1.4]$, where $f_\pi$ is the weak pion decay
constant. Indeed, the value given above for $a(M_\rho)$ in eq.(\ref{fita}) corresponds to
$Q=712$ MeV, a rather natural value for $Q$. This is important in order
to conclude that the physics encoded in this subtraction constant just
corresponds to re-scattering between the mesons, because if the subtraction
constant required  unnatural values of the cut-off then one should
conclude that some intrinsic dynamics is playing a role besides pure
meson-meson self-interactions.

In addition, we also consider the situation where we do not include either the
$\eta_8\eta_8$ state, like in ref.\cite{npa}, but allowing for different
values of the subtraction constants. Performing the fit we finally have:
\be
\label{fitb}
a_{\pi\pi}=-1.1,\,\, a_{K\bar{K}}=-1.6,\,\, a_{K\pi}=-0.8,\,\,
a_{K\eta}=-0.8\,\, a_{\pi\eta}=-0.5~. 
\ee
The resulting curves are the short-dashed ones of fig.\ref{fig:1s0kswcm}. 

We take these three T-matrices as reflecting the uncertainty
in our approach for the coupling constants and pole positions presented in
table \ref{table:coup}. The point is that the $\eta_8\eta_8$ channel plays
some role above 1 GeV, see for example the long-dashed line in the
$(1-\eta_0^2)/4$ panel, but on the other hand one can also reproduce the data
without this channel, short-dashed lines. Indeed, as already noted in ref.\cite{nd}, 
 the inelasticity for the $I=0$ S-wave 
partial wave, $\eta_0$, cannot be appropriately reproduced when the
$\eta_8\eta_8$ channel is taken into account unless 
a bare singlet pole around 1 GeV is included \cite{nd}, compare the
long-dashed and solid lines in fig.\ref{fig:1s0kswcm}. Thus, although the
preexisting resonances can also be avoided for energies lower than 1.2
GeV, they certainly give some contribution particularly in connection with he $\eta_8\eta_8$ channel. 
 We have then considered the
three situations presented above, where the $\eta_8\eta_8$ as well as the
preexisting resonances are included \cite{nd} (solid lines), where the
preexisting resonances are removed but the $\eta_8\eta_8$ is kept (long-dashed
lines) and finally when both the preexisting resonances and the $\eta_8\eta_8$
channel are removed (short-dashed lines). On the other hand, it is apparent
from fig.\ref{fig:1s0kswcm} that the bare poles are not negligible above 1.2 GeV.

\begin{figure}[ht]
\psfrag{Re}{Re$\sqrt{s}$ (MeV)}
\psfrag{Im}{Im$\sqrt{s}$ (MeV)}
\centerline{\epsfig{file=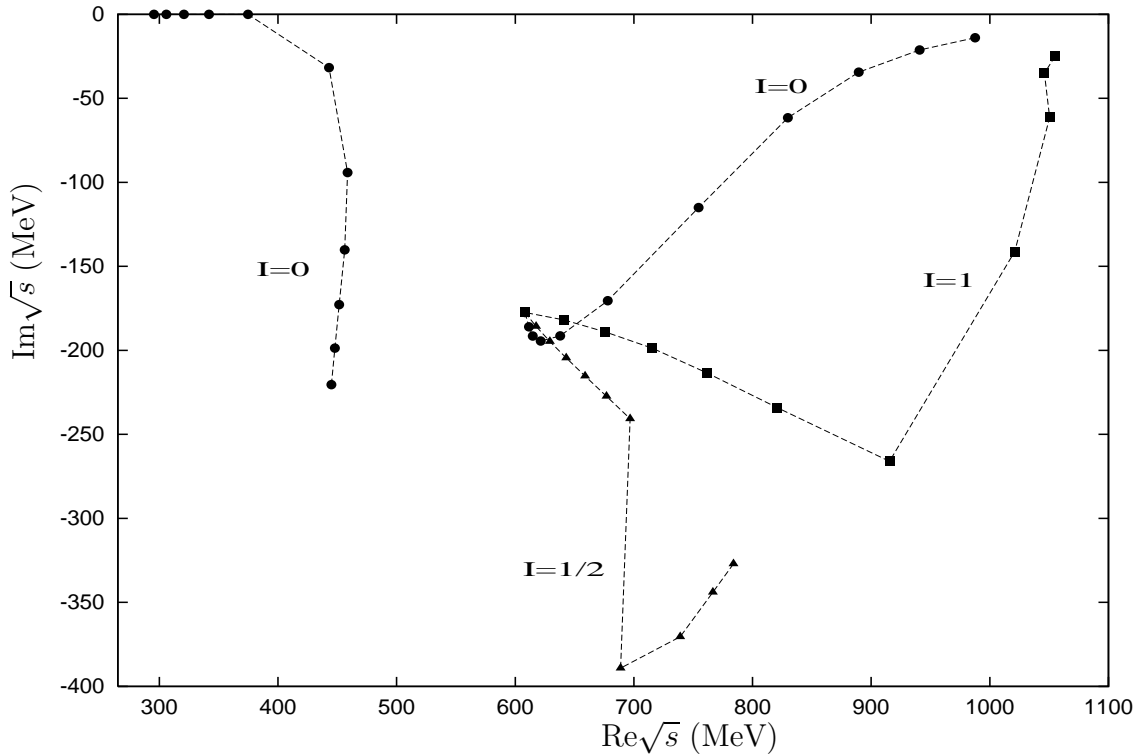,height=4.0in,width=6.0in,angle=0}}
\vspace{0.2cm}
\caption[pilf]{\protect \small Continuous movement of the octet and singlet poles from a SU(3) symmetric point 
to the physical limit. The different isospin channels are indicated in the figure and the step in the 
 $\lambda$ parameter is 0.1.
\label{fig:path}}
\end{figure}

We now perform a smooth transition from the physical limit to a SU(3)
symmetric one through a continuous parameter $\lambda \in [0,1]$, such that:
\ba
\label{smooth}
        m_\pi(\lambda)&=&m_\pi+\lambda(m_0-m_\pi) ~,\nn\\
        m_K(\lambda)&=&m_K+\lambda(m_0-m_K)~, \nn\\
        m_\eta(\lambda)&=&m_\eta+\lambda(m_0-m_\eta)~, \nn\\
        f_K(\lambda)&=&f_K+\lambda(f_\pi-f_K) ~,\nn \\
        f_\eta(\lambda)&=&f_\eta+\lambda(f_\pi-f_\eta)~.
\ea
So, for $\lambda=0$ one has the physical limit while for $\lambda=1$ one ends with a SU(3) 
symmetric point. The interesting point for us is to note the continuous movement of the
        pole locations of the $\kappa$, $a_0(980)$, $\sigma$ and $f_0(980)$ from their physical positions
 ($\lambda=0$) to a degenerate octet plus a singlet as shown in
        fig.\ref{fig:path}, for a common SU(3) mass $m_0=350$ MeV  and a step in $\lambda$
        of 0.1. This clearly shows that an octet and a singlet of 
resonances in the SU(3) limit evolve and give rise to the physical poles of the lightest scalars so 
that these resonances must be considered as the lightest nonet of scalar resonances. Thus, the $\kappa$ and 
$a_0(980)$ are pure octet states while the $\sigma$ and $f_0(980)$ are a mixture of the singlet 
and octet $I=0$ states.

We now perform a standard SU(3) analysis of the coupling constants of the lightest scalar resonances 
to pairs of pseudoscalars within the aforementioned SU(3) assignment. The coupling constants are 
obtained by determining the residues at the pole positions of the resonances through the expression:
\be
T_{ij}=\lim_{s\rightarrow s_R}-\frac{g_i g_j}{s-s_R}~,
\ee
where $s_R$ is the corresponding pole position and the subscripts $i$ and $j$ indicate the 
two pseudoscalar channels. 

Then, for the different options taken above to fit data we show in table
\ref{table:coup} the modules of the couplings since the phases can be 
contaminated from background contributions \cite{mp}.
\begin{table}[ht]
\begin{center}
\begin{tabular}{|l|r|r|r|r|}
\hline
Resonance & Full ref.\cite{nd} & No bare poles & No bare
poles & Average\\
 &  & eq.(\ref{fita})& eq.(\ref{fitb})& \\
 & GeV & GeV & GeV & GeV\\
\hline
\hline
$\sigma$         & $0.445-i\,0.220$ &   $0.443-i\,0.213$ & $0.442-i\,0.214$ &
$(0.443\pm 0.002)-i\,(0.216\pm0.004) $   \\ 
& $|g_{\pi\pi}|=3.01$& $|g_{\pi\pi}|=2.94$  & $|g_{\pi\pi}|=2.95$ &
$|g_{\pi\pi}|=2.97\pm 0.04 $      \\ 
& $|g_{K\bar{K}}|=1.09$& $|g_{K\bar{K}}|=1.30 $  & $|g_{K\bar{K}}|=1.34$ &
$|g_{K\bar{K}}|=1.24\pm 0.13$    \\
& $|g_{\eta_8\eta_8}|=0.09$  & $|g_{\eta_8\eta_8}|=0.04$ &  &
$|g_{\eta_8\eta_8}|=0.07\pm 0.04$\\
\hline 
$f_0(980)$ & $0.988-i\,0.014$ & $0.983-i\,0.007$  & $0.987-i\,0.011$ &
$(0.986\pm 0.003)-i\,(0.011\pm0.004)$ \\
         & $|g_{\pi\pi}|=1.33$ & $|g_{\pi\pi}|=0.89$  & $|g_{\pi\pi}|=1.18$ & $|g_{\pi\pi}|=1.1\pm0.2$    \\ 
         & $|g_{K\bar{K}}|=3.63$ & $|g_{K\bar{K}}|=3.59 $  &
         $|g_{K\bar{K}}|=3.83 $ & $|g_{K\bar{K}}|=3.68\pm 0.13$  \\
         & $|g_{\eta_8\eta_8}|=2.85$ & $|g_{\eta_8\eta_8}|=2.61$   & & $|g_{\eta_8\eta_8}|=2.73\pm$ 0.17\\

\hline 
$a_0(980)$ &  $1.055-i\,0.025$ &  $ 1.032-i\,0.042$  & $1.030-i\,0.086$ &
$(1.039\pm 0.014))-i\,(0.05 \pm 0.03)$ \\
&$|g_{\pi\eta_8}|=3.88$  & $|g_{\pi\eta_8}|=3.67$  &  $|g_{\pi\eta_8}|=4.08$&
$|g_{\pi\eta_8}|=3.9 \pm 0.2$ \\ 
        &   $|g_{K\bar{K}}|=5.50$ & $|g_{K\bar{K}}|= 5.39$  & $|g_{K\bar{K}}|=
        5.60$& 
$|g_{K\bar{K}}|= 5.50\pm 0.11$  \\
\hline 
$\kappa$ & $0.784-i\,0.327$ & $0.804-i\,0.285$  & $0.774-i\,0.338$& $(0.787\pm
0.015)-i\,(0.32\pm 0.03)$ \\
& $|g_{K\pi}|=5.02$ & $|g_{K\pi}|=4.93$  & $|g_{K\pi}|=4.89$ &
$|g_{K\pi}|=4.94\pm 0.07$ \\ 
        & $|g_{K\eta_8}|=3.10$& $|g_{K\eta_8}|=2.96 $  & $|g_{K\eta_8}|=3.00$ 
& $|g_{K\eta_8}|=3.02\pm 0.07$      \\
\hline
\end{tabular}
\caption{\protect \small Coupling constants for the scalar resonances for the solutions of ref.\cite{nd} and those 
corresponding to eqs.(\ref{fita}) and (\ref{fitb}). Let us remember that the $I=0$ 
states $|\pi\pi\rangle_0$ and $|\eta_8\eta_8\rangle_0$ of eq.(\ref{unor}) are normalized to 1/2 
and not to 1 so as to take care of the invariance of these states under the exchange of the two $\pi$ or
 $\eta_8$. In the fifth column we show the average of the values in the three previous columns.
\label{table:coup}}
\end{center}
\end{table}

Let us consider first the $a_0(980)$ and $\kappa$ resonances that do not have singlet contribution 
and should be considered as pure octet resonances.  Note that 
although one can also have an anti-symmetric octet representation, $8_a$, from the tensorial product 
$8\otimes 8$, is not allowed that a pair of pseudoscalars in the SU(3) limit stay in S-wave in 
this representation because of Bose-Einstein statistics. For the reader convenience, we write the 
scattering states with definite SU(3) eigenvalues for the symmetric octet, $8_s$, and the 
singlet, $1$:
\ba
\label{su3states}
|8_s, I=1, I_3=0\rangle&=&\frac{1}{\sqrt{5}}|\pi^0 \eta_8\rangle+\frac{1}{\sqrt{5}}|\eta_8 \pi^0º
\rangle+\sqrt{\frac{3}{20}}(|K^+K^-\rangle - |K^0 \bar{K}^0\rangle)
+\sqrt{\frac{3}{20}}(|K^- K^+\rangle - |\bar{K}^0 K^0\rangle)\nn\\
&=&\frac{1}{\sqrt{5}}|\pi\eta_8\rangle+\frac{1}{\sqrt{5}}|\eta_8\pi\rangle-\sqrt{\frac{3}{10}} 
|K\bar{K}\rangle_1-\sqrt{\frac{3}{10}}|\bar{K}K\rangle_1~,\nn\\
|8_s,I=\frac{1}{2},I_3=\frac{1}{2}\rangle&=&\sqrt{\frac{3}{20}}(|\pi^0 K^+\rangle + 
|K^+ \pi^0\rangle)+\sqrt{\frac{3}{10}}(|\pi^+ K^0\rangle+|K^0 \pi^+\rangle)
-\frac{1}{\sqrt{20}}(|K^+ \eta_8\rangle+|\eta_8 K^+\rangle) \nn\\
&=&-\frac{3}{\sqrt{20}}|\pi K\rangle_{1/2}+\frac{3}{\sqrt{20}}|K\pi\rangle_{1/2}-\frac{1}{\sqrt{20}}
|K\eta_8\rangle-\frac{1}{\sqrt{20}}|\eta_8 K\rangle~,\nn\\
|8_s,I=0,I_3=0\rangle&=&\frac{1}{\sqrt{5}}(|\pi^0\pi^0\rangle+|\pi^+\pi^-\rangle+
|\pi^-\pi^+\rangle)-\frac{1}{\sqrt{20}}
(|K^+K^-\rangle+|K^-K^+\rangle\nn\\
&+&|K^0\bar{K}^0\rangle+|\bar{K}^0 K^0\rangle)
-\frac{1}{\sqrt{5}}|\eta_8\eta_8\rangle\nn\\
&=&-\sqrt{\frac{6}{5}}|\pi\pi\rangle_0+\frac{1}{\sqrt{10}}|K\bar{K}\rangle_0
-\frac{1}{\sqrt{10}}|\bar{K} K\rangle_0-\sqrt{\frac{2}{5}}|\eta_8\eta_8\rangle_0~,\nn\\
|1,I=0,I_3=0\rangle&=&\frac{1}{\sqrt{8}}
(\pi^0\pi^0+\pi^+\pi^- +\pi^-\pi^+ +\eta_8\eta_8+K^+ K^-+K^-K^+ + K^0\bar{K}^0 + \bar{K}^0 K^0)\nn\\
&=&-\frac{\sqrt{3}}{2}|\pi\pi\rangle_0
+\frac{1}{2}|\eta_8\eta_8\rangle_0-\frac{1}{2}|K\bar{K}\rangle_0+\frac{1}{2}|\bar{K}K\rangle_0~.
\ea
Above we have also shown the decomposition of the  octet or singlet states in isospin 
states in addition to its decomposition in two pseudoscalar pairs. 
Furthermore, since the $|\pi\pi\rangle_0$ and $|\eta_8\eta_8\rangle$ are symmetric states 
under the exchange of the two pseudoscalars, we have added an extra $1/\sqrt{2}$ in the expression 
for the isospin states, so that:
\ba
\label{unor}
|\pi\pi\rangle_0&=&-\frac{1}{\sqrt{6}}|\pi^0\pi^0+\pi^+\pi^-+\pi^-\pi^+\rangle~, \nn\\
|\eta_8\eta_8\rangle_0&=&\frac{1}{\sqrt{2}}|\eta_8\eta_8\rangle~.
\ea

From this decomposition and identifying the $a_0(980)$ as the $|8_s,I=1,I_3=0\rangle$ state and 
the $\kappa^+$ as the $|8_s,I=1/2,I_3=1/2\rangle$ resonance, we have the following identities 
for their couplings:
\ba
\label{su3a0kappa}
g(a_0\rightarrow \pi\eta_8)&=&\frac{1}{\sqrt{5}}\, g_8~,\nn\\
g(a_0\rightarrow (K\bar{K})_1\rangle&=&-\sqrt{\frac{3}{10}}\, g_8~,\nn\\
g(\kappa \rightarrow (K\pi)_{1/2})&=&\frac{3}{\sqrt{20}}\, g_8~,\nn\\
g(\kappa \rightarrow K \eta_8)&=&-\frac{1}{\sqrt{20}}\, g_8~.
\ea
As a result, we predict the following three independent relations:
\ba
\frac{g(a_0\rightarrow \pi \eta_8)}{g(a_0\rightarrow (K\bar{K})_1)}&=&-\sqrt{\frac{2}{3}} = -0.82~,\nn\\
\frac{g(\kappa\rightarrow (K\pi)_{1/2})}{g(\kappa\rightarrow K\eta_8)}&=&-3~,\nn\\
\frac{g(a_0\rightarrow (K\bar{K})_1)}{g(\kappa\rightarrow(K\pi)_{1/2})}&=&-\sqrt{\frac{2}{3}} = 
-0.82~.
\ea

Averaging the values given in table \ref{table:coup} we obtain:
\ba
\label{coup12}
\left|\frac{g(a_0\rightarrow \pi\eta_8)}{g(a_0\rightarrow (K\bar{K})_1)}\right|&=&0.70 \pm 0.04~,\nn\\
\left|\frac{g(\kappa\rightarrow (K\pi)_{1/2})}{g(\kappa\rightarrow K\eta_8)}\right|&=&~1.64\pm 0.05,\nn\\
\left|\frac{g(a_0\rightarrow (K\bar{K})_1)}{g(\kappa\rightarrow(K\pi)_{1/2})}\right|&=&1.10\pm 0.03 ~.
\ea
For the first and third ratio the agreement is within the expected accuracy of around a $20\%$. 
It is worth stressing that the third ratio relates the couplings between both resonances and then is 
sensitive to the fact that $g_8$ is a common parameter in the SU(3) analysis. However, for the
 second ratio the disagreement is by a factor of two and hence for this particular 
coupling the SU(3) expectation does not hold within the expected accuracy. Nevertheless, we will show below that 
when  SU(3) breaking is allowed in the SU(3) analysis we recover the agreement
for the second ratio as well. Thus, these
results, together with the observed soft evolution of the poles from 
the SU(3) limit to the physical one, support the picture of the $a_0(980)$ and the $\kappa$ as members 
of the same octet. Still, we shall give below an additional analysis to support this conclusion. Notice 
that since the third ratio is in agreement with our analysis, the coupling $g(\kappa\rightarrow 
K\eta_8)$ is the one in disagreement with the present SU(3) expectations, 
as confirmed below when we allow SU(3) breaking linear in the quark masses.

 In 
addition to the ratios in eq.(\ref{coup12}) we can also obtain $|g_8|$ from the values of 
$|g(a_0\rightarrow \pi\eta_8)|$, $|g(a_0\rightarrow (K\bar{K})_1)|$ and $|g(\kappa\rightarrow (K\pi)_{1/2})|$
  in table \ref{table:coup} and the SU(3) relations in eq.(\ref{su3a0kappa}):
\be
\label{g8}
|g_8|=8.7\pm 1.3 \hbox{ GeV}~,
\ee
with a relative error around a $15\%$, well within the accuracy of an SU(3) analysis. In the number 
above we have not included the value of $|g(\kappa\rightarrow K\eta_8)|$ since, as discussed above,
 departs more than the allowed $20\%$  from the SU(3) expectations at this stage.

Let us now consider the $I=0$ resonances, namely, the $\sigma$ and the $f_0(980)$. These resonances 
are expressed as a mixture of the $I=0$ singlet and octet states, $|S_1\rangle$ and 
$|S_8\rangle$, respectively:
\ba
\label{relcoup}
|\sigma\rangle&=&\cos \theta |S_1\rangle +\sin \theta |S_8\rangle~,\nn\\
|f_0(980)\rangle &=& -\sin \theta |S_1\rangle +\cos \theta |S_8\rangle~,
\ea
with $\theta$ the mixing angle. Then we have the following relations for the couplings:
\ba
\label{sigf0coup}
g(\sigma\rightarrow (\pi\pi)_0)&=&-\frac{\sqrt{3}}{4} \cos \theta g_1-\sqrt{\frac{3}{10}} \sin \theta g_8 
~,\nn \\
g(\sigma \rightarrow (K\bar{K})_0)&=&-\frac{1}{2}\cos \theta g_1 +\frac{1}{\sqrt{10}} \sin \theta g_8~,\nn\\
g(\sigma \rightarrow (\eta_8\eta_8)_0)&=&\frac{1}{4}\cos \theta g_1-\frac{1}{\sqrt{10}}\sin \theta g_8~,
\nn\\
g(f_0\rightarrow (\pi\pi)_0)&=&\frac{\sqrt{3}}{4}\sin\theta g_1-\sqrt{\frac{3}{10}} \cos \theta g_8~,\nn\\
g(f_0\rightarrow (K\bar{K})_0)&=&\frac{1}{2}\sin\theta g_1+\frac{1}{\sqrt{10}}\cos\theta g_8~,\nn\\
g(f_0\rightarrow (\eta_8\eta_8)_0)&=&-\frac{1}{4}\sin\theta g_1-\frac{1}{\sqrt{10}}\cos\theta g_8~.
\ea
From table \ref{table:coup} there is a very stark result, the almost vanishing coupling, much 
smaller than any other, of the $\sigma$ resonance with the $\eta_8\eta_8$ channel. 
Imposing then that $g(\sigma\rightarrow \eta_8\eta_8)=0$ in the equations above we obtain the 
relation:
\be
\frac{g_1}{g_8}=\sqrt{\frac{8}{5}} \tan \theta~.
\label{g1g8}
\ee
Employing this result, it follows that:
\ba
\label{xy}
\frac{g(\sigma\rightarrow (\pi\pi)_0)}{g_8}&=&-2\sqrt{\frac{3}{10}} \sin\theta~,\nn\\
\frac{g(f_0\rightarrow (K\bar{K})_0)}{g_8}&=&\frac{1}{\sqrt{10}} \cos\theta(1+2\tan^2\theta)~.\nn
\ea
We take the previous equations to calculate simultaneously $g_8$ and $\theta$. Nevertheless, 
since we only know the modules of the different couplings, as shown in table \ref{table:coup}, we can 
only determine the modules of $\theta$ and $g_8$. Notice that from eq.(\ref{g1g8}), the relative sign 
between $g_1$ and $g_8$ is fixed by the chosen one of $\theta$. Once this is fixed, a change of sign in 
$g_8$ only gives rise to a global change of sign in the coupling constants listed in 
eq.(\ref{sigf0coup}) since from eq.(\ref{g1g8}) it implies a simultaneous change  in the sign of $g_1$ as 
well. Thus, it is a matter of convention to choose the sign of $g_8$ and $\theta$ with our information 
of the modules of the coupling constants given in table \ref{table:coup}. However, once the signs of 
$g_8$ and $\theta$ are chosen then from eq.(\ref{g1g8}) the sign of $g_1$ is fixed:
\ba
\label{coup}
|g_8|&=&10.5\pm0.5 \,\hbox{ GeV}~,\nn\\
\cos^2{\theta}&=&0.934\pm 0.006~,\nn\\
|\theta|&=&  14.9^o \pm 0.7^o~.
\ea

 Interestingly,  the consistency of our approach is notorious by comparing the value of $g_8$ 
from the study of the $a_0$ and $\kappa$ resonances with 
that in eq.(\ref{coup}) from the study of the mixing in the $I=0$ system. We 
see that both values agree within errors  which at most are a 15$\%$. We show as well the average 
value of $g_8$  from the equation above and  eq.(\ref{g8}), so that:
\be
\label{g82}
|g_8|=9.6\pm 1.3 \,\hbox{ GeV}.
\ee
From eq.(\ref{g1g8}), we have the following value for $g_1$:
\be
\label{singlet}
|g_1|=3.5\pm 0.5 \,\hbox{ GeV}~.\nn\\
\ee

Now we use a more versatile way of treating eqs.(\ref{su3a0kappa}) and (\ref{sigf0coup}) which is 
also more suited for the case when further SU(3) breaking is considered. For that
we employ the method of the least squares. We construct the $\chi^2$ function as:
\ba
\label{chi}
\chi^2=\sum_{R,PQ} \frac{|g(R\rightarrow PQ)-\hat{g}(R\rightarrow PQ)|^2}{\sigma_{R,PQ}^2}~,
\ea
where the sum extends over all the resonances $\sigma$, $f_0(980)$, $\kappa$, $a_0(980)$ and all the pseudoscalar 
pairs $PQ$ as shown in table \ref{table:coup}. We have denoted by $\hat{g}(R\rightarrow PQ)$ those 
coupling constants calculated from the SU(3) analysis, while $g(R\rightarrow PQ)$ refers to the 
average ones in table \ref{table:coup}.

In order to perform the numerical analysis based on the minimization of
eq.(\ref{chi}), we have allowed for different signs in the evaluation of
eqs.(\ref{sigf0coup}) since in table \ref{table:coup} we only know the modules of the coupling 
constants. The only consistent fit, where the ascribed signs for the
couplings in the minimization process and those that finally come up coincide,
is very similar to our previous results, eqs.(\ref{coup}), eq.(\ref{g82}) and (\ref{singlet}). 
The precise values are:
\ba
\label{xinobreak}
|g_1|=3.9\pm 0.8\,\hbox{ GeV} ~,\nn\\
|g_8|=8.2\pm 0.8\,\hbox{ GeV}~,\nn\\
|\theta|=19^o\pm 5^o~,
\ea
such that the relative sign of $g_1$ with respect to $g_8$ is in agreement 
with eq.(\ref{g1g8}). In order to calculate the errors we have multiplied by 
 a common factor the errors of the coupling constants in the fifth column of 
table \ref{table:coup}, much smaller than the expected accuracy of the
SU(3) analysis,  so that the $\chi^2$ per degree of freedom is one. Our 
results are clearly compatible with the previously obtained values for $g_8$ (\ref{g82}), 
$g_1$ (\ref{singlet}) and $\theta$ (\ref{coup}). The resulting couplings are:

\ba
\label{coup0su3}
|g(a_0\rightarrow \pi\eta_8)|&=&   3.7\pm 0.4 \,\hbox{ GeV}~,\nn\\
|g(a_0\rightarrow (K\bar{K})_1)|&=&4.5 \pm 0.4 \, \hbox{ GeV}~,\nn\\
|g(\kappa\rightarrow (K\pi)_{1/2}|&=& 5.5 \pm 0.5 \,\hbox{ GeV}~,\nn\\
|g(\kappa\rightarrow K\eta_8 )|&=&  1.8\pm 0.2 \,\hbox{ GeV}~,\nn\\
|g(\sigma\rightarrow (\pi\pi)_0)|&=&3.0 \pm 0.5 \, \hbox{ GeV}~,\nn\\
|g(\sigma\rightarrow (K\bar{K})_0)|&=& 0.9 \pm 0.5 \,\hbox{ GeV}~,\nn\\
|g(\sigma\rightarrow (\eta_8\eta_8)_0)|&=& 0.0\pm 0.3 \,\hbox{ GeV}~,\nn\\
|g(f_0\rightarrow (\pi\pi)_0)|&=& 3.7 \pm 0.5 \,\hbox{ GeV}~,\nn\\
|g(f_0\rightarrow (K\bar{K})_0)|&=& 3.1 \pm 0.3 \,\hbox{ GeV}~,\nn\\
|g(f_0\rightarrow (\eta_8\eta_8)_0)|&=&2.8 \pm 0.2 \, \hbox{ GeV}~.
\ea

We see a remarkable agreement between the calculated couplings from our SU(3)
analysis in eq.(\ref{coup0su3}) and the averages reported in the fifth column of
table \ref{table:coup}. Of the ten couplings there are only two
exceptions. The already discussed $g(\kappa\rightarrow K\eta_8)$ which turns
around a factor 2 too small and the $g(f_0\rightarrow (\pi\pi)_0)$ which is
too large by a factor around 3.  It should 
be noted that the $|g(f_0\rightarrow (\pi\pi)_0)|$ coupling is the most sensitive  to the presence or 
absence of a preexisting state in the $f_0(980)$ resonance since it changes by a 40$\%$, while for 
the other couplings the change is lower than a $5\%$. This points towards a special status for this 
coupling which controls the width of the $f_0(980)$ resonance. Indeed the large reduction, by a 
factor of 2, between 
the imaginary part of the pole position of the $f_0(980)$ resonance in the third
column of table \ref{table:coup} reflects the reduction 
in the value of its coupling to the $(\pi\pi)_0$ system. Notice that the ratio of the square of the 
couplings of the $f_0$ to $|\pi\pi\rangle_0$ given in the second and third columns is 
$(0.89/1.33)^2\simeq 1/2$, as corresponds to the ratio 
of the widths. In addition, as noticed in ref.\cite{isgur,npa}, the $f_0(980)$ behaves as a
 pure $K\bar{K}$ bound state when the 
$\pi\pi$ channel is removed. As a result, one expects that the $g(f_0\rightarrow (\pi\pi)_0)$ coupling is mostly
driven by kaonic loops decaying into two pions so that the SU(3) breaking can be
huge due to the large difference between the $K\bar{K}$ and $\pi\pi$ thresholds.

Summarizing, of the ten couplings constant  we have used three of them 
to calculate $\theta$, $g_8$, $g_1$, and then we 
end with the values given in eq.(\ref{coup0su3}). Between the latter, eight
couplings are in very good agreement with their 'experimental' values given in
table \ref{table:coup} but two of them are in disagreement, although the coupling 
$g(\kappa\rightarrow K\eta_8)$ will agree with our SU(3) analysis once further SU(3) breaking 
corrections are taken into account, eq.(\ref{su3lag}). On the other hand,
we have also checked that the obtained value 
of $g_8$ from the study of the pure octet states $\kappa$ and $a_0(980)$, eq.(\ref{g8}), is compatible 
with that from the study of the mixing in the $I=0$ system, eq.(\ref{coup}). The average 
of both values is the one quoted in eq.(\ref{g82}), compatible with the value
of $g_8$ given in eq.(\ref{xinobreak}) from the minimization of the $\chi^2$ function, 
eq.(\ref{chi}). This is an important check and then, as a 
whole picture, we can
 claim that the $\sigma$, $f_0(980)$, $a_0(980)$ and $\kappa$ constitute a nonet of dynamically generated
 scalar resonances with a mixing angle of around $20^o$. The preexisting bare
 pole contributing to the $f_0(980)$ resonance appears as a small contribution 
to the $f_0(980)$ physical pole that particularly affects the couplings of this resonance with the 
$\pi\pi$ channel, as apparent when comparing the couplings of the $f_0(980)$ in the second column 
of table \ref{table:coup} with those in the third one.

As stated above, we consider a second SU(3) coupling constant analysis to
support our view that these 
resonances constitute a nonet. We now consider the couplings shown in table \ref{table:eigencoup} 
of the $\sigma$, $f_0(980)$, 
$a_0(980)$ and $\kappa$ scalar resonances with the two pseudoscalar states of
eq.(\ref{su3states}) with well defined SU(3) eigenvalues.
\begin{table}
\begin{center}
\hspace{-0.5cm}
\begin{tabular}{|l|r|r|r|r|}
\hline
Resonance & Full ref.\cite{nd} & No bare poles  & No bare poles &Average \\
 &            & eq.(\ref{fita})   &  eq.(\ref{fitb}) & \\
 & \hbox{GeV} & \hbox{GeV} & GeV &\\
\hline
\hline 
$\sigma$ & $|g(\sigma\rightarrow 1)|=3.65 $ & $|g(\sigma\rightarrow 1)|=3.86$  &
 $|g(\sigma\rightarrow 1)|=3.89$ &  $|g(\sigma\rightarrow 1)|=3.80\pm 0.13$   \\ 
         & $|g(\sigma\rightarrow 8_s)|=2.67 $  & $|g(\sigma\rightarrow 8_s)|=2.38$ 
  & $|g(\sigma\rightarrow 8_s)|=2.38$ & $|g(\sigma\rightarrow 8_s)|=2.48\pm 0.18$ \\
\hline
$f_0(980)$ & $|g(f_0 \rightarrow 1)|= 5.35 $ &$|g(f_0 \rightarrow 1)|=5.07 $ 
 &$|g(f_0 \rightarrow 1)|=4.20 $  &$|g(f_0 \rightarrow 1)|=4.9\pm 0.6 $  \\  
      & $|g(f_0\rightarrow 8_s)|=4.15 $ & $|g(f_0\rightarrow 8_s)|= 3.90$ 
& $|g(f_0\rightarrow 8_s)|= 2.45$ & $|g(f_0\rightarrow 8_s)|= 3.5\pm 0.9$\\
\hline
$a_0$ & $|g_8|=8.95  $ & $|g_8|=8.83$  & $|g_8|=9.01$  & $|g_8|=8.5\pm 0.5$ \\
\hline 
$\kappa$ & $|g_8|=8.12 $ & $|g_8|=7.94$  & $|g_8|=7.90$  & $|g_8|=7.99\pm 0.12$   \\
\hline
\end{tabular}
\caption{\protect \small Coupling constants of the scalar resonances with two mesons states with well defined SU(3) 
eigenvalues.  In the fifth column we show the average of the values of the three previous columns.
\label{table:eigencoup}}
\end{center}
\end{table}

Since the $a_0(980)$ and $\kappa$ are pure octet resonances, their couplings to an octet 
state must be the same and this is indeed what happens as can be seen from the values given 
in table \ref{table:eigencoup}, which are the same  within an error of around 9$\%$. 
Averaging the values of $g_8$ in the first three columns of table \ref{table:eigencoup}, we have:
\be
\label{g83}
|g_8|=8.5\pm 0.5 \, \hbox{ GeV}~,
\ee 
a very similar value to that of eq.(\ref{xinobreak}). The weighted value of $g_8$  between 
the ones given in eqs.(\ref{g82}) and (\ref{g83}) is:
\be
\label{g8final}
|g_8|=8.6 \pm 0.5 \,\hbox{ GeV}~.
\ee
Because the $f_0$ and $\sigma$ are a mixture between the pure singlet and the $I=0$ octet states, 
eq.(\ref{relcoup}), their couplings to the singlet and octet states are driven
 by this mixing and then we have:
\ba
g(\sigma\rightarrow 1)&=&\cos \theta \,g_1~,\nn\\
g(\sigma\rightarrow 8_s)&=& \sin \theta \,g_8~,\nn\\
g(f_0\rightarrow 1)&=&-\sin  \theta \,g_1 ~,\nn\\
g(f_0\rightarrow 8_s)&=&\cos  \theta \,g_8~.
\label{leading}
\ea
Taking into account the second line of the previous equations and the calculated  
value of $g_8$ in eq.(\ref{g8final}), we then have a new calculation for
$g_1$ and  the mixing angle $\theta$. 
\ba
\label{angle}
|g_1|&=&3.96\pm 0.13\, \hbox{ GeV}~,\nn\\
\cos^2\theta&=&0.916\pm0.006~,\nn\\
|\theta|&=&16.8^o\pm 0.7^o~,
\ea 
very similar to the values of eq.(\ref{xinobreak}) and compatible within the shown errors from 
the dispersion in the values shown in tables \ref{table:coup} and \ref{table:eigencoup}. 

Let us notice that we can also find a solution for the last two equations in (\ref{leading}). 
One then obtains very different values to those of eqs.(\ref{xinobreak}) 
and (\ref{angle}) with $\theta\simeq 65^0$, for $g_8$ given in eq.(\ref{g83}). 
Indeed it is easy to see by finding $\tan\theta$ from either of the ratios 
$g(\sigma\rightarrow 8_s)/g(f_0\rightarrow 8_s)$ or $g(f_0 \rightarrow 1)/g(\sigma\rightarrow 1)$ 
that the set of four equations in (\ref{leading}) is not compatible with the values given
in table \ref{table:eigencoup}. This is why we have solved separately those two
equations of the $\sigma$, whose solution is given in eq.(\ref{angle}), and those of the $f_0(980)$. 
Indeed, the $\chi^2$ (\ref{chi}) calculated from the set of values of eq.(\ref{angle}) is 1.5 smaller
 than that  calculated from the values that result if we considered instead the 
$f_0$ couplings in eqs.(\ref{leading}). Furthermore, if we take the latter as starting
values to minimize the $\chi^2$ function,  we end then with a solution similar
to eqs.(\ref{xinobreak}) and (\ref{angle}). Let us note from
eq.(\ref{su3states})  that the 
two pion system enters both in the octet and singlet $I=0$ states with large SU(3) Clebsch-Gordan 
coefficients and then this last method cannot be applied to the  $f_0(980)$ since 
it was not good enough for the study of the $g(f_0\rightarrow (\pi\pi)_0)$ coupling. It is also 
worth stressing that the problem with the $g(f_0\rightarrow (\pi\pi)_0)$ coupling cannot be solved by considering an extra singlet contribution to the $f_0(980)$ resonance, as one could think because of the preexisting singlet contribution found in ref.\cite{nd} to the physical $f_0(980)$. This is so because the
 $g(f_0\rightarrow 8_s)$ coupling does not involve any singlet contribution and is in disagreement with
 our expectations from eq.(\ref{leading}), with the values determined in eqs.(\ref{g83}) and 
(\ref{angle}) from the sudy of the couplings of the $\sigma$, $a_0(980)$ and $\kappa$ resonances in table \ref{table:eigencoup}.

As state above we now consider SU(3) breaking in the couplings of the octet
and singlet states to two pseudoscalars. Since the singlet and the symmetric octet 
SU(3) representations are symmetric under the exchange of the two pseudoscalars,
at linear order in the quark mass matrix, we can write the following Lagrangian:
\be
\label{su3lag}
F\langle S(2\Phi \chi \Phi+\Phi^2\chi+\chi\Phi^2) \rangle+G S_1 \langle 2\Phi
\chi \Phi+\Phi^2\chi+\chi\Phi^2\rangle~.
\ee
The same expression can be deduced from the chiral Lagrangians of
ref.\cite{pich}. In the expression above $\Phi$ is the usual $8\times 8$
matrix with the octet of pseudoscalars and the same
applies for $S$ but now referring to the scalar octet. For more details see e.g. ref.\cite{pich}. 
Finally, $S_1$ is the
scalar singlet, $\chi=diag(0,0,m_K^2-m_\pi^2)\propto(0,0,m_s-\hat{m})$ is the
matrix responsible for the SU(3) breaking with  $m_s$ the strange quark mass and $\hat{m}$ 
the average of the masses of the two lightest quarks. On the other hand, $F$ and $G$ are two 
constants. We have then minimized the $\chi^2$, eq.(\ref{chi}), with
the corrections derived from the previous 
Lagrangian added to eqs.(\ref{su3a0kappa}) and
(\ref{sigf0coup}). We want 
to emphasize here that the values of the couplings $g_8$, $g_1$ and $\theta$ turn out to be the same and the 
SU(3) breaking parameters $F (m_K^2-m_\pi^2)=-0.23\pm0.16$ and $G (m_K^2-m_\pi^2)=0.13\pm 0.4$ take 
very natural values, around a $20\%$ of unity as expected from SU(3) breaking. 
The main novelty is that now $g(\kappa\rightarrow K\eta_8)=2.52$, quite close
to the value given in table \ref{table:coup}. Nevertheless, the situation for the $g(f_0\rightarrow (\pi\pi)_0)$
coupling does not improve and we are still a factor 3 off. For completeness,
we also discuss here a qualitative different fit to that of
eq.(\ref{xinobreak}) where the SU(3) breaking is much larger but the $\chi^2$
function is smaller, by around a factor of 2, than that
corresponding to the fit in eq.(\ref{xinobreak}) but including now the just mentioned SU(3) breaking
corrections. This fit arises when allowing a change of sign
in the coupling $g(\kappa\rightarrow K\eta_8)$  with respect to the case 
without the SU(3) breaking from eq.(\ref{su3lag}). Let us note that
this coupling is SU(3) suppressed because of the smallness of the
corresponding Clebsch-Gordan, see eq.(\ref{su3states}), so that moderate SU(3)
corrections could well change its sign. However, in this case the
SU(3) breaking parameters are larger by around a factor of 7 to those given
above, so that $F (m_K^2-m_\pi^2)=1.44\pm 0.11$ and $G
(m_K^2-6m_\pi^2)=0.81\pm 0.2$ and hence they drive to large SU(3) corrections
that are difficult to accept. We see e.g. that the resulting $g_8$ parameter
is lower by a factor of two than that obtained directly from the $a_0(980)$
and $\kappa$ analysis in eq.(\ref{g83}), by directly averaging the values shown in table
\ref{table:eigencoup}. This is of course very unlikely to be accepted since
we see a nice agreement between the couplings of the $a_0(980)$ and $\kappa$
 to the pure octet states in table \ref{table:eigencoup} as corresponds to a
 soft SU(3) breaking. Furthermore, from this fit to the two pseudoscalar
 couplings we can then predict the values of the couplings of the resonances
 to the SU(3) eigenstates in eq.(\ref{su3states}) and then we observe that
 the predicted value for $g_8$ is much smaller, by around a factor 20, than 
the ones shown in table \ref{table:eigencoup}. As a result we discard this fit as
 non-physical and keep the one of eq.(\ref{xinobreak}) as our final results.

By passing, we mention that taking the real part of the poles presented in table
\ref{table:coup} as the masses of the corresponding resonances, we then find
that they fulfill in good agreement a {\it linear}
Gell-Mann-Okubo mass relation with the mixing angle given in eq.(\ref{xinobreak}):
\be
4m_\kappa-m_{a_0}=3(m_{f_0} \cos^2 \theta+m_\sigma \sin^2 \theta)~.
\ee
The ratio between the difference of the right(2.7) and left(2.1) sides of
the above equation and the average of both is around a $25\%$, well shaped
within the usual SU(3) accurateness. On the other hand, the
quadratic Gell-Mann-Okubo mass relation is much worse behaved, being the
relative difference around a factor 2.

Finally, the sign of $\theta$ can be fixed by considering the strength of the scalar form factors:
\ba
\label{sff} 
f_n&=&\langle 0|\frac{1}{\sqrt{2}}(\bar{u}u+\bar{d}d)|K\bar{K} \rangle_0~,\nn\\
f_s&=&\langle 0|\bar{s}s|K\bar{K} \rangle_0~,\nn\\
\ea
at the peak of the $f_0(980)$. In the previous equations we have denoted by $|0\rangle$ the vacuum state. 
These form factors, among others, were calculated in ref.\cite{meisso}, 
and the peak of the $f_0(980)$ is clearly visible. Now, from ref.\cite{meisso} at the $f_0(980)$ peak
 the ratio $|f_s/f_n|\simeq 5.5$, 
denoting clearly that the $f_0(980)$ couples stronger with strangeness,
as well known. Notice that the 
coupling of the $f_0(980)$ to the two kaons factorizes out in the ratio and
then one is only sensitive to the $f_0$ coupling to the quark-antiquark sources. We can perform
the SU(3) decomposition of the previous sources:
\ba
\bar{s}s&=&\frac{1}{\sqrt{3}} s_1-\sqrt{\frac{2}{3}} s_8~,\nn\\
\bar{n}n \equiv\frac{1}{\sqrt{2}}(\bar{u}u+\bar{d}d)&=&\sqrt{\frac{2}{3}}
s_1+\frac{1}{\sqrt{3}} s_8~,
\ea
where $s_1$ and $s_8$ are singlet and octet operators. Then, neglecting any
other $SU(3)$ breaking beyond mixing we can write:
\ba
\label{ffsresults}
\langle 0|\bar{s}s |f_0(980)\rangle&=&-\sin \theta\frac{1}{\sqrt{3}}
\langle 0|s_1|S_1\rangle-\cos\theta \sqrt{\frac{2}{3}}\langle 0|s_8|S_8\rangle~,\nn\\
\langle 0|\bar{n}n |f_0(980)\rangle&=&-\sin \theta \sqrt{\frac{2}{3}}\langle
0|s_1|S_1 \rangle+\cos\theta \frac{1}{\sqrt{3}}\langle 0|s_8|S_8\rangle~.
\ea
Now, from the value $|\theta|=19^o$ given in eq.(\ref{xinobreak}) and taking,
just for an order of magnitude estimation, 
$\langle 0|s_1|S_1\rangle \simeq \langle 0|s_8|S_8\rangle $, as required by 
U(3) symmetry, we then have:
\ba
\theta=+19^o&\rightarrow& \left|\frac{f_s}{f_n}\right|=3.5~,\nn\\
\theta=-19^o&\rightarrow& \left|\frac{f_s}{f_n}\right|=0.7~.
\ea
Thus, the solution with $\theta>0$ is clearly favored since requires much 
smaller U(3) breaking corrections than that with $\theta<0$, which would be huge. 
As a result, if we take the convention that $g_8>0$,   we can remove
the absolute value signs in eq.(\ref{xinobreak}), and write:
\ba
\label{definite}
g_8&=& 8.2 \pm 0.8 \, \hbox{ GeV}~,\nn\\
g_1&=& 3.9 \pm 0.8\, \hbox{ GeV} ~,\nn\\
\theta&=&+19^o\pm 5^o
\ea
It is also worth mention that our mixing angle is in agreement with the determination from 
ref.\cite{nap}.

\section{Conclusions}
\label{sec:conc}
We have first shown that the $\sigma$, $f_0(980)$, $a_0(980)$ and $\kappa$ poles evolve
continuously in the $\lambda$ parameter, eq.(\ref{smooth}), from the physical limit $\lambda=0$
 to a nonet of resonances in a SU(3) limit case, $\lambda=1$, made up by a degenerate octet 
and a singlet. Indeed we already showed this in ref.\cite{nd}, but now we have analysed 
in further detail the situation where the bare poles, an octet with a mass around 1.4 GeV and 
a singlet with a mass around 1 GeV, are removed.
In particular, we have now performed two fits to the scattering data of the S-waves in the $I=0$, 1/2 and 1 
channels without bare resonances and with a common value for all the subtractions constants, 
eq.(\ref{fita}), or with different ones and without the $\eta_8\eta_8$ channel as well, eq.(\ref{fitb}). 
In addition, 
we have also interpreted that a value around $-1$ for the subtraction
constants at the scale $\mu=M_\rho$ corresponds to calculating the unitarity
loops with a three-momentum cut-off of a natural value, $Q\simeq M_\rho$. 
Although this certainly shows that the aforementioned resonances constitute a
nonet of dynamically generated scalar 
resonances with the $a_0(980)$ and $\kappa$ being members only of the octet, we have performed 
two additional analyses that have shown: First, the SU(3) limit picture 
as a whole is still useful in order to study the lightest scalar nonet in the physical limit 
(within the expected precision of around a 20$\%$ for such estimations). Second, we consistently 
conclude again that those resonances belong to the same nonet and third, we are able to 
obtain a value for the mixing angle between the $I=0$ states of the singlet
and octet multiplets, together with the coupling constants 
$g_8$ and $g_1$. In this way we consistently study within SU(3) symmetry, 
in terms of three parameters, 
the set of couplings: $g(a_0\rightarrow 
(K\bar{K})_1)$, $g(a_0\rightarrow \pi\eta_8)$, $g(a_0\rightarrow 8_s)$, 
$g(\kappa\rightarrow (K\pi)_{1/2})$, $g(\kappa\rightarrow K\eta_8)$, 
$g(\kappa\rightarrow 8_s)$, $g(\sigma\rightarrow (\pi\pi)_0)$
, $g(\sigma\rightarrow (K\bar{K})_0)$, $g(\sigma\rightarrow (\eta_8\eta_8)_0)$, 
$g(\sigma \rightarrow 1)$, $g(\sigma\rightarrow 8_s)$, $g(f_0\rightarrow (K\bar{K})_0)$, 
$g(f_0\rightarrow (\eta_8\eta_8)_0)$. We have seen that the coupling
$g(\kappa\rightarrow K\eta_8)$ is finally in agreement with our SU(3) analysis
once linear SU(3) breaking in the quark masses is taken into account from 
eq.(\ref{su3lag}). Nevertheless, the coupling $g(f_0(980)\rightarrow
(\pi\pi)_0)$ require still 
higher orders in the SU(3) breaking parameters and do not 
follow the standard simple scheme developed here in the lines of the well established vector and 
tensor nonets. The given SU(3) analyses take the values of the couplings of the 
scalar resonances with two pseudoscalar states with either well defined isospin or well 
defined SU(3) eigenvalues. Both  analyses have different sensitivity to
the SU(3) breaking, since different combinations 
of amplitudes are involved, and follow very different numerical analyses,
but remarkably, compatible values between errors for the mixing angle as well
as for the coupling $g_8$ and $g_1$ stem. This consistency is worth stressing
and gives us further 
confidence in our results together with our interpretation that the failure in
reproducing the $(\pi\pi)_0$ coupling of the $f_0(980)$ should be considered as an isolated
fact in an overall good description and interpretation of table
\ref{table:coup}. We have also established that the scalar resonances fulfill
to good precision a linear Gell-Mann-Okubo mass relation with our determined
mixing angle and masses in table \ref{table:coup}. Finally, from a study of
the scalar form factors calculated in ref.\cite{meisso}, we have fixed the
sign of the mixing angle to be positive and then we conclude that $\theta=+19^o\pm 5^o$.

\section*{Acknowledgments}
I would like to acknowledge useful discussions with D. Black and E. Torrente. This work is partially
supported by the DGICYT project FPA2002-03265.

\end{document}